\documentclass[pre,twocolumn]{revtex4}
\usepackage{amsmath,amssymb}
\usepackage{graphicx}

\newcommand{\href}[1]{}
\newcommand{\trans}{^{\text{T}}}
\newcommand{\R}{{\bf R}}
\newcommand{\x}{{\bf x}}
\newcommand{\z}{{\bf z}}
\newcommand{\y}{{\bf y}}

\renewcommand{\k}{{\bf k}}
\renewcommand{\d}{\mbox{d}}
\newcommand{\llangle}{\left\langle}
\newcommand{\rrangle}{\right\rangle}
\newcommand{\Id}{\operatorname{Id}}
\newcommand{\xcm}{\x_\text{cm}}
\newcommand{\bxcm}{\bar{\x}_\text{cm}}
\newcommand{\bxmm}{\bar{\x}_{m\mu}}

\newcommand{\bycm}{\bar{\y}_\text{cm}}
\renewcommand{\tt}{{t+\tau}}
\renewcommand{\r}{{\bf r}}
\newcommand{\A}{{\bf A}}
\newcommand{\B}{{\bf B}}
\newcommand{\C}{{\bf C}}
\newcommand{\q}{{\bf q}}

\begin{document}
\title{Intramolecular Fluorescence Correlation Spectroscopy in a Feedback Tracking Microscope}
\author{Kevin McHale} 
\altaffiliation{Laboratory of Chemical Physics, NIDDK, NIH, Bethesda, MD 20892}
\email{mchalek@gmail.com}
\author{Hideo Mabuchi}
\affiliation{Edward L. Ginzton Laboratory, Stanford University, Stanford, CA 94305}
\begin{abstract} 
We derive the statistics of the signals generated by shape fluctuations of large molecules studied by feedback tracking microscopy. We account for the influence of intramolecular dynamics on the response of the tracking system, and derive a general expression for the fluorescence autocorrelation function that applies when those dynamics are linear. We show that tracking provides enhanced sensitivity to translational diffusion, molecular size, heterogeneity and long time-scale decays in comparison to traditional fluorescence correlation spectroscopy. We demonstrate our approach by using a three-dimensional tracking microscope to study genomic $\lambda$-phage DNA molecules with various fluorescence label configurations.
\end{abstract}
\maketitle
\section{Introduction}
Fluorescence Correlation Spectroscopy (FCS) is a method for optically measuring local concentration fluctuations of fluorescence-labeled molecules in solution\citep{Magde72a,Elson74a,Maiti97a,Krichevsky02a}. The fluorescence autocorrelation function --- referred to as the FCS curve --- contains signatures of the dynamic properties of those molecules, such as diffusion coefficients or reaction rates, which are inferred by comparison of the curve to theoretical predictions. Modern approaches to FCS use tightly focused laser beams and confocal detection to probe diffraction-limited sample volumes, are sensitive enough to measure fluorescence from single dye-labeled molecules, and can resolve fluctuations on time-scales as fast as the fluorescence lifetimes of the dyes\citep{Berglund02a,Nettels07a}. These methods have found success in a  wide range of applications in biology and chemistry.

One relatively new application of FCS is the study of the intramolecular dynamics of large polymer chains. Such motions were first described theoretically over 50 years ago\citep{Rouse53a, Zimm56a}, but were only coarsely probed experimentally because of the insensitivity of the experimental methods available at the time. As first demonstrated in 2003, FCS is sensitive to the internal motions of polymers that are large relative to the focused waist of the probe laser\citep{Lumma03a}. These initial measurements suggested that the internal dynamics of large double-stranded DNA (dsDNA) molecules are dominated by the stiffness of the polymer chain and by hydrodynamic couplings between spatially proximate polymer segments. Later measurements conflicted with these conclusions\citep{Shusterman04a}, however, and the resulting controversy has yet to be fully resolved despite several iterations of experimental and theoretical improvements\citep{Winkler06a,Petrov06a,Cohen07a}.

In a recent paper\citep{McHale09DNA}, we argue that FCS is not a sufficiently sensitive technique for characterizing the internal dynamics of large polymer molecules because the theoretical FCS curves contain too many free parameters, resulting in underdetermined numerical fits to the data. We instead used a technique based on feedback tracking microscopy, a technology that has been developed by our and other groups over the past few years\citep{Enderlein00a, Levi03a, Berglund04a, Andersson05a, Cohen05a, Lessard06a, Armani06a, Cang06a} (and reviewed recently in \citep{Cang08a}) by which a feedback system tracks the translational motion of a single molecule and keeps it in the focus of the microscope. In the particular variant of feedback microscopy developed by our group, we compute the autocorrelation function of the fluorescence measured from the tracked molecule and analyze it in a manner analogous to traditional FCS\citep{Berglund05a,Berglund07a,Berglund07b}. This tracking-FCS (tFCS) approach generates statistics that are related to traditional FCS (which we will refer to as \emph{stationary} FCS for clarity in this paper), but provides enhanced sensitivity to both translational and intramolecular motion. Our tFCS measurements revealed that the translational statistics and radius of gyration of large dsDNA molecules were together consistent with strong hydrodynamic interactions, as commonly expected, but that  the intramolecular relaxation statistics are surprisingly not consistent with the Zimm polymer model\citep{Zimm56a,Doi86a,McHale09DNA}. 

This paper provides the mathematical foundation for intramolecular tracking-FCS.  In section \ref{Section:GeneralTheory} we compute the tFCS curve for a molecule exhibiting conformation fluctuations. We account for the effect of those fluctuations on the response of the tracking system, as well as the systematic artifacts that the tracking system adds to the tFCS curve. We focus in particular on molecules described by linear dynamical models, both because these include the standard polymer dynamics theories and because it is possible to derive closed-form expressions for the tFCS curve for such models. We illustrate characteristics of the intramolecular tFCS approach with a simple example model for the molecular motion. In Section \ref{Section:Measurements} we demonstrate the application of tFCS to the study of the intramolecular motion of fluorescence-labeled genomic $\lambda$-phage DNA using a variation on the apparatus we described in \citep{McHale07a}. Our measurements are consistent with predictions for molecules with three different label configurations; furthermore, they reveal sensitivity to heterogeneity between molecules with random labeling schemes with a noise floor consistent with the predicted photon-counting noise level.% We conclude in Section \ref{Section:Related} with a brief comparison to related techniques, addressing dynamic light scattering and stationary FCS as well as image correlation analysis in an electro-osmotic trap, a technique similar to tFCS but that generates distinct statistics, as we will show.

\section{The intramolecular tracking-FCS curve}\label{Section:GeneralTheory}
\begin{figure}
\centering
\includegraphics[width=3.25in]{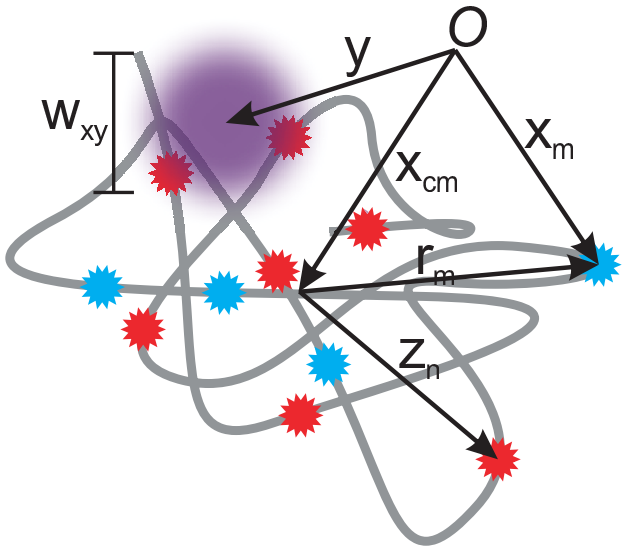}
\caption{Typical experimental scenario. The gray curve represents a molecule sparsely labeled with red and cyan dyes. The circular violet region represents the tFCS probe beam (the larger tracking beam is not illustrated). Vectors $\xcm$, $\x_m$ and $\y$ relative to the arbitrary origin $O$ represent the center of mass of the molecule, position of tFCS dye $m$, and beam position, respectively. Dye positions relative to $\xcm$ ($\r_m$ for tFCS dyes and $\z_n$ for tracking dyes) are used in the text for convenience.}\label{Fig:vectors}
\end{figure}

We begin with a description of the experimental scenario we are concerned with, illustrated in Figure \ref{Fig:vectors}. A molecule is labeled with a collection of two types of fluorescent dyes, distinguished in the figure by their colors. The red dyes are excited by the tracking laser to determine the position of the molecule, and the cyan dyes are excited by a probe laser for intramolecular tFCS measurements. The tracking system reacts to the motion of the molecule by adjusting the position of the laser beams to follow the center of mass of the tracking dyes. Fundamental localization noise induces tracking errors so the beams do not follow the target position precisely, as emphasized by their off-center displacement in the figure. The tFCS dyes emit fluorescence bursts whenever the intramolecular motion causes them to drift through the probe laser beam. The goal of this section is to calculate the statistics of this tFCS signal as determined by the dynamics of the tracking system and the statistics of the molecule's motion. 

We begin with the standard definition of the FCS curve, 
\begin{equation}
g_2(\tau) = \frac{\llangle I(t)I(\tt)\rrangle}{\llangle I(t) \rrangle\llangle I(\tt)\rrangle} - 1,
\label{eq:FCSDefn}
\end{equation}
where $\langle \cdot \rangle$ denotes an average over time, equivalent to an ensemble average due to the ergodicity of the fluorescence signal. In this paper we deal with $g_2(\tau)$ for a collection of $M$ dyes moving along a set of trajectories $\{\x_m^t\}$, each with its own possibly distinct dynamics (we will use this superscript notation as shorthand to indicate the time dependence of various quantities throughout this paper). We define the brightness $b_m$ of dye $m$ so that $b_m \phi_0$ is the fluorescence rate we detect when that dye is excited by a laser with intensity $\phi_0$. We can then write $I(t)$ in terms of the dye positions and brightnesses and the spatially-varying laser intensity $\phi(\x)$:
\begin{equation}
I(t) = \sum_{m=1}^M b_m \phi(\y^t - \x_m^t),
\label{eq:FluorIntensity}
\end{equation}
where $\y^t$ is the position of the laser beam over time. In the case where $\y^t$ is constant, this $I(t)$ is exactly that of stationary FCS.

We express Eq. \ref{eq:FluorIntensity} in terms of the Fourier transform of $\phi(\x)$, $\tilde \phi(\k)$, in order to facilitate calculations. The unnormalized autocorrelation $G(\tau) \equiv \llangle I(t)I(\tt)\rrangle$ is then given by
\begin{equation}
 G(\tau) = \sum_{m, \mu=1}^M b_m b_\mu \int \frac{\d^6 \bar \k}{(2\pi)^6} \llangle e^{i \bar \k\trans \left(\bxmm - \bar \y \right)}\rrangle \Phi(\bar \k),
\label{eq:FluorCorr}
\end{equation}
where we have simplified notation by defining the concatenated vectors $\bar \k = \begin{pmatrix}\k\\\k'\end{pmatrix}$, $\bar \y = \begin{pmatrix}\y^\tt \\ \y^t\end{pmatrix}$ and $\bxmm = \begin{pmatrix}\x_\mu^\tt \\ \x_m^t\end{pmatrix}$, and the product of the beam profiles $\Phi(\bar \k) = \tilde \phi(-\k) \tilde \phi(-\k')$. A similar expression exists for the time average $\langle I(t) \rangle$.

Equations \ref{eq:FCSDefn} and \ref{eq:FluorCorr} together define $g_2(\tau)$ in terms of the properties of the tracked molecule and tracking system; all that remains is to insert appropriate models for these dynamics. In the general case Eq. \ref{eq:FluorCorr} cannot be simplified because $\bxmm$ and $\bar \y$ are highly correlated. The tracking system responds to both the center-of-mass and intramolecular motion of the molecule, and these motions may be correlated with each other. For example, a molecule's shape fluctuations may couple into variations in its apparent translational diffusion coefficient\citep{Cohen07a}, and spatial inhomogeneities within the sample may cause the intramolecular dynamics to depend on the center-of-mass position. As a consequence, the only general approach to calculating the tFCS curve relies on solving for the joint statistics of the laser and the dyes from a set of coupled equations of motion. 

\subsection{Linear models in homogeneous samples}
Linear models such as the Rouse and Zimm polymer models predict Gaussian statistics for the molecular motion because the Brownian force is Gaussian. Our feedback system is linear as well, so it exhibits Gaussian statistics in response to the molecular motion\citep{Berglund07b}. If we assume that $\bxmm$ and $\bar \y$ are jointly Gaussian random variables, we can make a fairly dramatic simplification to the average in Eq. \ref{eq:FluorCorr} (see supporting information):
\begin{equation}
\llangle e^{i\bar\k\trans \left(\bxmm-\bar \y\right)}\rrangle\\ = \exp\Biggl\{-\frac 1 2 \bar\k\trans \llangle \left(\bxmm - \bar \y\right)^2\rrangle \bar\k\Biggr\}
\label{eq:GaussFullSimplified},
\end{equation}
where we have adopted the notation for the outer product $\x^2 \equiv \x\x\trans$ and have assumed that $\bxmm - \bar \y$ has zero mean, a condition that is enforced by the tracking system.  This simplification is helpful because it separates the average into second-order correlation matrices that are relatively easy to calculate. 

Linear models further predict that the center of mass motion and intramolecular motion are uncorrelated, provided the absence of any spatially inhomogeneous fields. These models express the dynamics using an eigenfunction expansion in which the center-of-mass motion appears as the zeroth-order mode, while the intramolecular motion is superposed in the higher-order modes. If we write $\x_m^t$ as a sum of the center-of-mass position $\xcm^t$ and the dye position relative to the center of mass $\r_m^t$, then the autocorrelation function is just the sum of autocorrelations. For example,
\begin{equation}
\llangle \x_\mu^\tt \left(\x_m^t\right)\trans\rrangle = \llangle \xcm^\tt \left(\xcm^t\right)\trans\rrangle + \llangle \r_\mu^\tt \left(\r_m^t\right)\trans\rrangle.
\end{equation}

Similarly, we can write $\y^t$ as a sum over components that respond to the center-of-mass motion, $\y_\text{cm}^t$; to the intramolecular motion, which we denote $\y_z^t$; and to localization noise arising from photon counting as described in \citep{Berglund06a}, which we denote $\y_F^t$. These three components are also uncorrelated: the first two because of the assumption of independence between center-of-mass and intramolecular motion, and the third because the photon counting process is independent of dye position. 

We now insert these decomposed terms into Eq. \ref{eq:GaussFullSimplified} to get
\begin{equation}
\llangle e^{i\bar\k\trans\left(\bxmm - \bar \y\right)} \rrangle = \exp\Biggl\{-\frac 1 2 \bar\k\trans \Bigl[ \Lambda_{m\mu}^\tau + \Sigma^\tau \Bigr] \bar \k \Biggl\},
\label{eq:GaussIndepSimplified}
\end{equation}
where we have defined
\begin{align}
\label{eq:LambdaIMDefn}
\Lambda_{m\mu}^\tau &= \Bigl\langle \left(\bar\r_{m\mu} - \bar \y_z\right)^2 \Bigr\rangle\\
\Sigma^\tau &= \llangle \left(\bxcm - \bycm\right)^2\rrangle + \llangle \left(\bar\y_F\right)^2\rrangle,
\label{eq:SigmaCMDefn}
\end{align}
and in those definitions we have used the $\bar \cdot$ notation to indicate concatenated two-time vectors as in Eq. \ref{eq:FluorCorr}. Defined in this way, $\Sigma^\tau$ constitutes a tracking error arising from finite system response bandwidth and from localization noise and is identical to that previously described in \citep{Berglund07b}. The intramolecular motion therefore adds a new time-dependent variance term to the systematic tracking error variance that was characterized previously. 

We now approximate the excitation laser beam with a three-dimensional Gaussian, as is standard practice in the FCS literature. We denote the covariance of this Gaussian by the diagonal matrix $\frac 1 4 W$, where the elements of $W$ --- denoted $w_x^2$, $w_y^2$ and $w_z^2$ --- are the squares of the beam waists along the x-, y-, and z-axes. Combining this with Eq. \ref{eq:GaussIndepSimplified}, we can compute the integrals in Eq. \ref{eq:FluorCorr} to write the FCS curve in terms of the matrices $\Sigma^\tau$ and $\Lambda_{m\mu}^\tau$:
\begin{equation}
g_2(\tau) + 1 \propto
\sum_{m, \mu = 1}^M b_m b_\mu \left\{\det\left[\Lambda_{m\mu}^\tau + \Sigma^\tau + \frac 1 4 \begin{pmatrix}W&0\\0&W\end{pmatrix}\right]\right\}^{-1/2},
\label{eq:FCSInsertable}
\end{equation}
where the constant of proportionality can by found by direct computation of $\langle I(t) \rangle$, or more simply by requiring that $g_2(\infty) = 0$. 

Equation \ref{eq:FCSInsertable} is suitable for any intramolecular model with linear dynamics. In general there may be coupling between the intramolecular motion along different Cartesian axes; for example, dye movement can induce fluid flows in the solvent that act upon adjacent dyes in a manner that must be described by a mobility tensor. However, the models that we will work with in this paper do not contain this feature. In this case, we may write the matrices $\Lambda_{m\mu}^\tau$ and $\Sigma^\tau$ in terms of smaller, diagonal matrices (denoting the $3\times 3$ identity by $\Id_3$):
\begin{align}
\Lambda_{m\mu}^\tau &= \begin{pmatrix}\lambda^0_{mm} \Id_3 & \lambda^\tau_{m\mu}\Id_3\\ \lambda^\tau_{\mu m} \Id_3& \lambda^0_{\mu\mu}\Id_3\end{pmatrix}
\label{eq:LambdaDecomp}\\
\Sigma^\tau &= \begin{pmatrix}\sigma^0 \Id_3 &\sigma^\tau\Id_3\\\sigma^\tau\Id_3&\sigma^0\Id_3\end{pmatrix}.
\label{eq:SigmaTauBreakup}
\end{align} 
For example, $\lambda^\tau_{m\mu} = \frac 1 3 \llangle \left(\r_\mu^\tt - \y_z^\tt\right)\trans\left(\r_m^t - \y_z^t\right)\rrangle$.
The quantities $\lambda^\tau_{m\mu}$ and $\sigma^\tau$ must then be found from the models chosen for the intramolecular and tracking system dynamics. We must note that, in addition to requiring independence, when expressed in this way $\Lambda_{m\mu}^\tau$ and $\Sigma^\tau$ also require that the dynamics are identical along all three Cartesian axes. In practice this idealization is rarely realized because the xy-axes and z-axis usually do not perform identically in our apparatus. We will therefore distinguish between the dynamics along different axes, when necessary, using the notation $\left(\lambda_{m\mu}^\tau\right)_x$ and $\sigma_x^\tau$ to indicate the x-axis, and likewise for the y- and z-axes. 

We may now write the tFCS curve in the most general form we will require in this paper:
\begin{widetext}
\begin{equation}
g_2(\tau) = \sum_{m,\mu=1}^M \frac{b_m b_\mu\left\{\sum_{m=1}^M b_m \prod_{\alpha\in\{x,y,z\}}\Bigl[\bigl(\lambda^0_{mm}\bigr)_\alpha + \frac 1 4 w_\alpha^2 + \sigma_\alpha^0\Bigr]^{-1/2}\right\}^{-2}}{\prod_{\alpha \in \{x,y,z\}}\left\{\Bigl[\bigl(\lambda^0_{mm}\bigr)_\alpha + \frac 1 4 w_\alpha^2 + \sigma_\alpha^0\Bigr]\left[\left(\lambda^0_{\mu\mu}\right)_\alpha + \frac 1 4 w_\alpha^2 + \sigma_\alpha^0\right] - \left[\left(\lambda^\tau_{m\mu}\right)_\alpha + \sigma_\alpha^\tau\right]^2\right\}^{1/2}} - 1,
\label{eq:GeneralFCSCurve}
\end{equation}
\end{widetext}
which reduces determination of $g_2(\tau)$ to the problem of finding $\lambda_{m\mu}^\tau$ and $\sigma^\tau$ for the particular models chosen for the intramolecular dynamics and the tracking system. In the next section, we compute the system response to the motion of the tracked molecule in order both to fully calculate $\sigma^\tau$ and to derive the necessary equations for calculation of the $\y_z$-dependent terms in $\lambda_{m\mu}^\tau$.

\subsection{Tracking system dynamics}\label{Section:TrackingSystem}
Earlier work by our group described the tracking stage as a linear control system that responds continuously to differences in position between it and the tracked particle, and furthermore characterized the fluorescence fluctuations that arise due to imperfect tracking fidelity\citep{Berglund07b}. Here we expand upon this work by incorporating the effect of intramolecular shape fluctuations on the tracking statistics. The approach taken here differs significantly from previous work in that we use a Langevin equation, rather than a Fokker-Planck equation, to compute the statistics we require. The advantage of this approach is that it is simpler to incorporate generalized intramolecular dynamics as a stochastic input to a deterministic tracking system than by finding and solving a Fokker-Planck equation that describes the joint molecular and tracking stage statistics. 

We let the molecule be labeled by a set of $N$ dyes to which the tracking system responds; these may be the same as those used to compute the FCS curve, but need not be. Therefore we must be somewhat careful in distinguishing between tracking dyes and FCS dyes. We let the tracking dyes move along a set of trajectories that we write as $\{\xcm^t + \z_n^t\}$, where $\z_n^t$ are positions relative to the center of mass that fluctuate according to the molecule's internal dynamics. As with $\xcm^t$ and $\r_m^t$ previously, we assume that $\xcm^t$ and $\z_n^t$ are uncorrelated. 

Provided that the molecule is smaller than the rotation radius and axial modulation distance of the tracking laser beam, the fluorescence from any individual dye on it will produce a localization signal that is proportional to the dye's distance from the tracking fixed point\citep{Berglund05a,McHaleThesis}. These signals add linearly for multiple dyes, so that the tracking system follows the average position of the $N$ tracking dyes.

We let $y^t$ be the position of the tracking stage along the x-axis, without loss of generality, as a function of time. Similarly, $x_\text{cm}^t$ and $z_n^t$ are the x-axis components of their corresponding vectors. We write a linear dynamical system to describe the evolution of $y^t$ in response to these dye positions and to a localization noise $F^t$ arising from photon counting as described in \citep{Berglund06a}: 
\begin{equation}
\begin{split}
\frac \d {\d t} \q^t &= \A \q^t + \B \left(x_\text{cm}^t + \frac 1 N \sum_{n=1}^N z_n^t + F^t\right)\\
y^t &= \C \q^t,
\label{eq:GenericDynamicalSystem}
\end{split}
\end{equation}
where $\q^t$ is an internal state variable and the matrices $\A$, $\B$, and $\C$ are a state-space realization of the tracking system dynamics. This abstract matrix formalism is used because the dynamics of the tracking system depend on those of all of its constituent components and, particularly with mechanical stages like our piezoelectric one, some of the system components may have nontrivial responses within the operational feedback bandwidth. While we can sometimes obtain satisfactory results with a simple first-order system (in which $\A$, $\B$, and $\C$ are scalars), often we require a two-dimensional system in order to account for the finite bandwidth of the mechanical stage\citep{Berglund07b}.

We now solve Eq. \ref{eq:GenericDynamicalSystem} in terms of its inputs, 
\begin{equation}
y^t = \C e^{\A t}\q^0 + \C \int_0^t \d \xi\, e^{\A(t-\xi)} \B \left(x_\text{cm}^\xi + \frac 1 N \sum_{n=1}^N z_n^\xi + F^\xi\right).
\label{eq:ExactSolForY}
\end{equation}
The first term in this equation is a transient one reflecting the state of the system at the initiation of tracking; we avoid consideration of such transients because the linearity assumption for the localization signal is often violated, and the resulting nonlinear statistics are quite complicated\citep{McHaleThesis}. Fortunately, the stability of the tracking system requires $\A < 0$, so that the transients decay sharply for $t$ larger than the time constant set by $|\A|^{-1}$.

We will use Eq. \ref{eq:ExactSolForY} in the steady-state limit $|\A|t \gg 1$ to compute the statistics of $y^t$. This limit can only be interpreted when $\z_n^t$ is stationary, but this will essentially always be the case. We assume that the center of mass moves by ordinary Brownian motion. The inputs $z_n^t$ and $F^t$ both have mean zero, so that $\llangle y^t \rrangle = x_\text{cm}^0$. For simplicity, we define our coordinate system so that $x_\text{cm}^0 = 0$.

The two-time correlation functions are more substantial. We break $y^t$ into its uncorrelated constituent parts, as described earlier: $y^t \equiv y_\text{cm}^t + y_z^t + y_F^t$. For each part we write out the product, insert the appropriate correlations, compute the integrals and take the steady-state limit. Beginning with the center-of-mass motion, we insert the Brownian correlation $\llangle x_\text{cm}^{t_1} x_\text{cm}^{t_2} \rrangle = 2D\min\{t_1, t_2\}$ and find, after some manipulation,
\begin{multline}
\llangle y_\text{cm}^\tt y_\text{cm}^t \rrangle = 2Dt - 2D\C\A^{-1}\\\times  \Bigl[\A^{-1}\left(\Id + e^{\A\tau}\right)\Gamma^\infty + \Gamma^\infty (\A\trans)^{-1}\Bigr]\C\trans,
\label{eq:CMResponse}
\end{multline}
where $\Gamma^\infty = \lim_{t\rightarrow \infty}\int_0^t \d\xi\, e^{\A(t-\xi)} \B\B\trans e^{\A\trans(t-\xi)}$ as defined in \citep{Berglund07b}. $\Gamma^\infty$ may alternatively be expressed as the solution of the equation $\A\Gamma^\infty + \Gamma^\infty \A\trans = -\B\B\trans$, which we exploit to simplify expressions containing this term. The simplification to Eq. \ref{eq:CMResponse} also requires $\C\A^{-1}\B = -1$, which indicates that the tracking system has no deterministic steady-state error and is guaranteed by the use of an integrating controller in the feedback loop\citep{Berglund07b}. The $\tau$-independent term in Eq. \ref{eq:CMResponse} represents the lag between the molecule and stage positions resulting from finite feedback bandwidth. As $|\A| \rightarrow \infty$ the lag term goes to zero because the tracking system follows the molecule with perfect fidelity.

Next we assume that $F^t$ is a white noise process with power spectral density $f$. Following the same procedure as for $y_\text{cm}^t$ yields
\begin{equation}
\llangle y_F^\tt y_F^t\rrangle = f^2 \C e^{\A\tau}\Gamma^\infty \C\trans
\label{eq:NoiseResponse}
\end{equation}
and puts us in position to compute $\sigma^\tau$ as defined in Eq. \ref{eq:SigmaTauBreakup}. We combine Eqs. \ref{eq:CMResponse} and \ref{eq:NoiseResponse} to get
\begin{equation}
\sigma^\tau = \C e^{\A\tau}\left[2D \A^{-1}\Gamma^\infty \left(\A\trans\right)^{-1} + f^2\Gamma^\infty \right]\C\trans,
\label{eq:SigmaTauGeneral}
\end{equation}
which is independent of $t$ because all $2Dt$ terms stemming from the center-of-mass motion have been canceled. This is the mathematical statement of the fact that the tracking system cancels the molecule's center-of-mass motion on time-scales longer than the tracking and intramolecular relaxation times. 

We cannot compute the component of the stage motion due to $z_n^t$ until we specify the model that we will use for the intramolecular dynamics. Next we present a simple example model that that captures the essential features of intramolecular tracking-FCS.

\subsubsection{Example: independent harmonically bound dyes}
We consider the example of a collection of dyes bound by a harmonic potential to a central point that is undergoing Brownian motion. Despite being somewhat artificial, this example will contain all of the essential details of intramolecular tracking-FCS and is convenient in that it is exactly solvable.

We let all dyes move independently of each other and with the same dynamics, and we require that motion along different Cartesian axes is uncorrelated. The dyes' intramolecular correlation function is a well-known result from the Ornstein-Uhlenbeck theory\citep{vanKampen92},
\begin{equation}
\llangle \r_\mu^\tt \left(\r_m^t\right)\trans\rrangle = e^{-\beta \tau} a^2 \delta_{m\mu}\Id_3,
\label{eq:OrnsteinUhlenbeck}
\end{equation}
where $\beta$ is the stiffness of the attractive bond to the central point and $a^2$ is the variance of the dye position. The distribution of these dyes is Gaussian, so the formula for $g_2(\tau)$ in Eq. \ref{eq:FCSInsertable} applies.

For simplicity, we choose a first-order dynamical model for our tracking system, with $\A\equiv-\gamma$, $\B = \gamma$ and $\C=1$. We may now compute the intramolecular response matrix elements $\lambda^\tau_{m\mu}$. If the sets of FCS dyes and tracking dyes are distinct, we have
\begin{equation}
\lambda_{m\mu}^\tau = a^2 e^{-\beta \tau} \delta_{m\mu} + \frac {a^2\gamma} N \left[\frac{\gamma e^{-\beta\tau}-\beta e^{-\gamma\tau}}{\gamma^2-\beta^2}\right].
\label{eq:harmonicDiff}
\end{equation}
The first term in this expression is the intramolecular correlation function, Eq. \ref{eq:OrnsteinUhlenbeck}. The second term arises because tracking errors add fluctuations to the fluorescence signal; it is strictly positive, indicating that in this configuration the fluorescence fluctuations are always larger than if the tracking system had an exact estimate of the center-of-mass position (as in the limit $N \rightarrow \infty$). In the particular case where the same set of dyes are used both as tracking indicators and tFCS probes, the result is
\begin{equation}
\lambda_{m\mu}^\tau = a^2 e^{-\beta \tau} \delta_{m\mu} - \frac {a^2\gamma} M \left[\frac{\gamma e^{-\beta\tau}-\beta e^{-\gamma\tau}}{\gamma^2-\beta^2}\right].
\label{eq:harmonicSame}
\end{equation}
Here the tracking error component is strictly negative because the fluctuations in the dye position relative to the probe beam are suppressed by the tracking system.

\begin{figure}%
\centering
\includegraphics[width=3.25in]{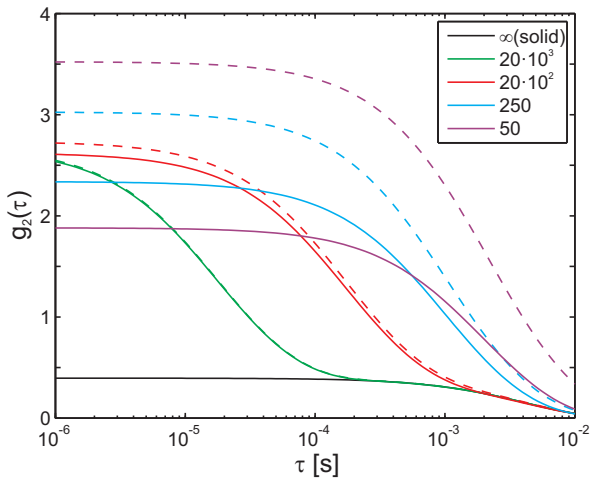}%
\caption{Example tracking-FCS curves for the harmonic model. Dashed curves indicate distinct tracking and probe dyes (Eq. \ref{eq:harmonicDiff}), and solid curves indicate identical tracking and probe dyes (Eq. \ref{eq:harmonicSame}).  The legend indicates the value of $\beta$ for each curve. All curves used $N=M=1$, $a=200$nm, $\sigma^0=($250nm$)^2$ for all axes, $w_{xy}=280$nm, and $w_z=800$nm.}%
\label{Fig:ExtFCS}%
\end{figure}

Figure \ref{Fig:ExtFCS} illustrates the tFCS curves for the harmonic model with varied values for $\beta$, including the limit $\beta \rightarrow \infty$ representing a solid particle (this limit does not account for rotational motion, so the particle it represents is densely labeled). For $\beta \gg \gamma$, the time-scales of the intramolecular motion and of the tracking system response are separated and the curves for the harmonic models approach that of the solid particle for $\tau > \beta^{-1}$. In this case, the molecule's rapid internal fluctuations average away on time-scales relevant to the tracking system. The tracking system is able to follow the intramolecular motion for smaller $\beta$; consequently, the fluorescence fluctuations increase significantly when the tracking and probe dyes are distinct and decrease significantly when the dyes are identical.

In the next section, we use the harmonic model to demonstrate features of the tFCS approach that generalize qualitatively to other models of molecular motion.
\subsection{General features of intramolecular tracking-FCS}
Tracking-FCS shares many general properties with stationary FCS. This section is not a survey of these familiar properties, but instead points out differences between the two approaches or features that are unique to tFCS. For a general survey of stationary FCS, see \citep{Krichevsky02a}. We also include a more detailed discussion of the differences between stationary FCS, tFCS and related image correlation techniques in the supporting information.
\subsubsection{Center of mass statistics and tracking errors}\label{Section:MSD}
The independent measurement of the position of the tracked molecule over time is a hallmark difference between stationary FCS and tracking techniques, allowing us to make very accurate diffusion coefficient measurements while simultaneously greatly suppressing the fluorescence decay due to translational motion. This is covered extensively in \citep{Berglund07a,Berglund07b} for when the tracked molecule is a point particle; here we examine the effect that intramolecular motion has on the center-of-mass statistics.

The most basic characterization of the performance of a particle tracking system is the size of its center-of-mass tracking error, which we usually define for each axis by $\sigma^0$, the variance of the stationary distribution of the quantity $y^t - \xcm^t$. Using the harmonically bound dye model with first-order tracking dynamics as an example, we find
\begin{equation}
\operatorname{Var}\left[ y^t - x_\text{cm}^t\right] = \frac{D}{\gamma}+\frac {f^2\gamma}2+\frac{a^2\gamma}{N(\gamma+\beta)},
\end{equation}
where the first two terms are familiar from point particle tFCS and the third results from the response of the stage to the intramolecular motion. Thus the intramolecular term adds a steady-state error that, when $N$ is small and $\beta \lesssim \gamma$, can actually dominate the tracking system's response. In some cases this may be desirable: for instance, if we were tracking the diffusion of a single point on a molecule as it moves within the molecule itself. In such cases it may be more sensible to define the tracking error differently, because $\xcm$ is no longer the relevant target position; the derivations in this paper would facilitate doing so.

\begin{figure}%
\centering
\includegraphics[width=3.25in]{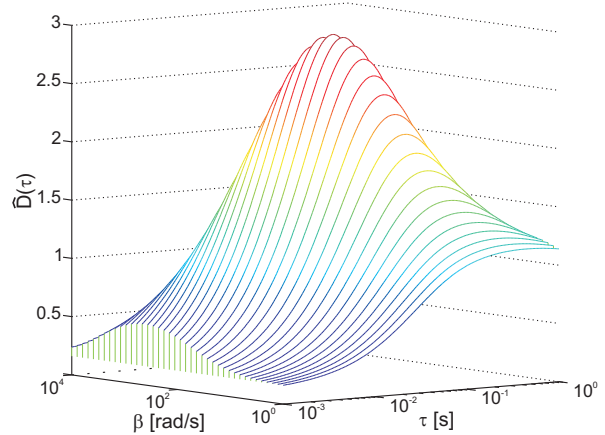}%
\caption{$\hat D(\tau)$, as defined in the text, evaluated for the harmonic model over a range of values for $\beta$. We used $D=1\mu$m$^2$/s, $a=500$nm, $\gamma=10$Hz, $f=8$nm/$\sqrt{\text{Hz}}$, $N=1$.}%
\label{Fig:MSDharmonic}%
\end{figure}
While the tracking error is important as a figure of merit, a more general and practically useful quantity is the variance of the stage increment, 
%\begin{multline}
%\operatorname{Var}\left[y^\tt - y^t\right] = 2D\tau - \frac{2D}{\gamma}\left(1-\frac{f^2\gamma^2}{2D}\right)\left(1-e^{-\gamma\tau}\right)\\
%+ \frac{2a^2\gamma^2}{N(\gamma^2-\beta^2)}\left[\frac{\gamma-\beta}{\gamma} - e^{-\beta \tau} + \frac \beta \gamma e^{-\gamma\tau}\right],
%\label{eq:MSDharmonic}
%\end{multline}
to which we fit curves in order to determine $\gamma$, $f$ and $D$ from our data. Figure \ref{Fig:MSDharmonic} illustrates the effect of the intramolecular motion on the quantity $\hat D(\tau) \equiv \operatorname{Var}\left[y^\tt - y^t\right](2\tau)^{-1}$, defined so that $\hat D(\infty) = D$. The sigmoid-like shape of the curve for small $\beta$ is what is typically observed due to pure center-of-mass tracking\citep{Berglund07b}. The curve peaks as a function of both $\tau$ and $\beta$ due to the tracking system's response to the intramolecular motion. The peak along $\tau$ indicates the trade-off between intramolecular motion on short time-scales and center-of-mass motion on long time-scales. The peak along $\beta$ occurs because slow intramolecular fluctuations (small $\beta$) are dominated by center-of-mass motion while fast intramolecular fluctuations (large $\beta$) are not tracked well due to latency in the feedback loop.
\subsubsection{Concentration and molecular size}
An obvious practical difference between stationary FCS and tFCS is that the tracking approach is not directly sensitive to the sample concentration. The concentration determines how frequently molecules drift into focus, but once a single molecule is detected and tracked its fluctuations are independent of the number of molecules elsewhere in the sample. By contrast, the sample concentration has a large influence in stationary FCS, determining the variance of the fluorescence fluctuations and consequently the overall scaling of the FCS curve through the value $g_2(0)$. 

The fluorescence variance is somewhat more complicated in tFCS, as it is partially determined by tracking errors (exclusively so in point-particle tFCS). Unlike in stationary FCS, however, the fluorescence variance in tFCS also contains a component due to intramolecular fluctuations. We demonstrate this using the harmonic model as an example, and for simplicity we assume that the tracking errors are small and unaffected by intramolecular motion, i.e. $\sigma^0 \ll a^2$ and $N \rightarrow \infty$. In the limit that the molecule is much bigger than the beam waists, i.e. $a \gg w/2$, the variance of the FCS curve is
\begin{equation}
 g_2(0) \approx \frac 1 M \frac{2^{3/2}a^3}{w_{xy}^2w_z}\equiv \frac 1 {\bar C \bar V},
\label{eq:BigMoleculeEffConc}
\end{equation}
where we have defined the effective intramolecular dye concentration $\bar C = M/a^3$ and imaging volume $\bar V = 2^{-3/2}w_{xy}^2w_z$. Equation \ref{eq:BigMoleculeEffConc} has exactly the same form as in stationary FCS except with the sample concentration replaced by its tFCS analog, $\bar C$.

It is straightforward to show that the tFCS variance is dominated by tracking errors in the small-particle limit $a \ll w/2$, where it approaches its value for point particles. Hence intramolecular tFCS has the same general property as intramolecular stationary FCS, in which the FCS curve is dominated by intramolecular motion for molecules larger than the excitation beam and by translational motion for smaller molecules\citep{Lumma03a}. An important difference, however, is that $g_2(0)$ is affected --- strongly so for large particles --- by molecular size in tFCS whereas it is determined by the sample concentration alone in stationary FCS\citep{Winkler06a}. This fact makes tFCS more sensitive to molecule size than stationary FCS and improves the numerical conditioning of fits to $g_2(\tau)$.
\subsubsection{Sensitivity to heterogeneity}
Stationary FCS is a true single-molecule method in the sense that it generates fluorescence signals from only one molecule at a time. However, the signal from a single molecule is never sufficient to determine a detailed FCS curve because only a small number of photons are typically detected from each molecule, so that statistical counting errors are very large. This is quite unlike tFCS, in which long observation times with the tracked molecule located in the brightest part of the excitation beam can enhance the signal-to-noise ratio by a factor of 100 or more\citep{McHale07a}. Tracking-FCS can therefore be used to determine complete FCS curves on individual molecules over a wide range of time-scales, providing the unique ability to resolve differences in the dynamics of different molecules. We will demonstrate this in our measurements in the next section.

\subsubsection{Sensitivity to decays on long time-scales}\label{Section:LongTime}
Another characteristic feature of tFCS is sensitivity to fluorescence decays on time-scales much longer than the characteristic diffusion time of a molecule through the laser focus. $g_2(\tau)$ only decays to zero when the quantities $\lambda_{m\mu}^\tau$ and $\sigma^\tau$ are both approximately zero; this means that decays in the intramolecular term $\lambda_{m\mu}^\tau$ are detectable even if they occur at times much longer than the center-of-mass diffusion time. Stationary FCS, by contrast, contains a decay of the form $g_2(\tau) \propto \tau^{-1}$ for $\tau$ longer than the characteristic diffusion time\citep{Krichevsky02a}, so that any longer decays are sharply attenuated.

\section{Application to double-stranded DNA}\label{Section:Measurements}
We now apply the theory developed in the previous sections to measurements on double-stranded DNA. 

\subsection{Linear polymer models}
The simplest dynamical model for the motion of a flexible polymer chain was developed by Rouse\citep{Rouse53a, Doi86a}, and forms the basis for subsequent refinements by Zimm\citep{Zimm56a} and others of the physics incorporated into the models. The Rouse model describes the polymer as a sequence of beads, connected and held together by springs and driven by independent Brownian forces. The discrete set of equations of motion for these beads is transformed into a partial differential equation for the polymer backbone by taking a continuum limit, assuming that the RMS distance between beads is much smaller than the overall length $L$ of the molecule. The resulting equation defines a time-dependent space curve $\R(u, t)$ parameterized by the position $u \in [0, L]$ along the backbone contour, and is solved using a Fourier series expansion. From this solution we find the correlation function needed to compute $g_2(\tau)$\citep{Doi86a},
\begin{multline}
\llangle \R(u, \tt) \R(u', t)\trans \rrangle = 2Dt\Id_3\\
+\frac{2r_0^2}{\pi^2} \sum_{q=1}^\infty \frac 1 {q^2} e^{-\tau/\tau_q}\cos\left[\frac{\pi q u}L\right]\cos\left[\frac{\pi q u'}L\right]\Id_3,
\label{eq:RouseCorrFcn}
\end{multline}
where $r_0^2$ is the mean squared distance between the end and center of mass of the polymer and $\tau_q$ is the relaxation time of Fourier mode $q$, which scales as $q^{-2}$ for the Rouse model and approximately as $q^{-3/2}$ for the Zimm model. We provide $\lambda_{m\mu}$ for a polymer with probe dyes at contour positions $\{u_m\}$ and tracking dyes at $\{v_n\}$ in Eq. S2 in the supporting information.

We have already demonstrated that the Rouse model is sufficient to describe tFCS measurements on $\lambda$-phage DNA, and that in fact the Zimm model is not consistent with our data\citep{McHale09DNA}. We will therefore only use the Rouse model in our fits in this paper. The results here are intended to illustrate the theoretical consistency and capabilities of the tFCS technique and the technical aspects of measurements on labeled polymers. We focus more on the specific quantities that tFCS is sensitive to than on the scientific interpretation of those quantities although, together with our closely-related work in \citep{McHale09DNA}, the technical developments described in this paper have provided a significant advance in the field of experimental DNA dynamics.
\subsection{Labeling}
The simplest way to incorporate fluorescence labels into DNA is with intercalating dyes that insert themselves randomly at sites approximately uniformly distributed over the DNA backbone. It is possible to incorporate many of these dyes, which makes them valuable for tracking since the molecule becomes brightly fluorescent with little influence of intramolecular motion on the dynamics of the tracking system. Intercalating dyes are less useful for the probe dyes which, in contrast to the tracking dyes, should be incorporated quite sparsely in order to maximize sensitivity to intramolecular motion. The difficulty is that the dye positions are random and are different for each molecule; both tFCS and stationary FCS are sensitive to these differences, but not sufficiently sensitive to infer  the dye label configurations precisely. This configuration uncertainty must be properly addressed when analyzing tFCS and stationary FCS data, and inevitably reduces the sensitivity of these techniques to the underlying polymer dynamics.

An alternative labeling procedure uses methods from molecular biology to incorporate dyes site-specifically into the DNA backbone. This is more challenging, but provides much greater experimental sensitivity because the dye configurations are exactly known and are identical for all molecules.

We prepared three different DNA-dye conjugates for our experiments: one molecule with intercalating tracking and probe labels; one molecule with intercalating tracking labels and a single probe label on the molecule's terminus; and one molecule with individual tracking and probe labels on opposite termini. We provide theoretical formulae for the tFCS curves for each of these molecules, derived from Eq. \ref{eq:RouseCorrFcn}, in the supporting information.

\subsection{Materials and methods}
We made all measurements using the three-dimensional tracking microscope described in \citep{McHale07a}, which we enhanced as described in \citep{McHale09DNA} by the addition of a second excitation laser at 444nm and a confocal detection channel with peak sensitivity at 480nm. 

We purchased genomic $\lambda$-phage DNA (48502 bp) from NEB and Invitrogen and produced three different DNA-dye conjugates. One conjugate was labeled with the intercalating dyes POPO-3 (for tracking) and POPO-1 (for tFCS), both purchased from Invitrogen. Molecules with mean inter-dye spacings of 300 base pairs (bp) for the tracking dye and between 300 and 48000 bp for the tFCS dye were prepared by adding DNA to dilute solutions of dye in TE buffer (10mM TRIS, 1mM Na$_2$EDTA, pH 8.0) and incubating at room temperature for 20 minutes. At these relatively low dye densities, we anticipate little alteration of the DNA dynamics\citep{Lumma03a}.

One conjugate was labeled by incorporation of a single Atto-425 conjugated dATP (Jena Biosciences) into a terminal single-stranded overhang using Klenow exo- DNA polymerase (NEB). The reaction was cleaned up using the QIAEX II silica adsorption procedure (Qiagen) and repeated ultrafiltration in a Microcon YM-100 unit (Millipore) until no free dyes were detected in solution. This conjugate was then labeled with POPO-3 for tracking as described above.

A final conjugate was prepared by ligating the biotinylated oligonucleotide sequence 5'-GGGCGGCGACCT-3'-Bio (IDT) onto the free single-stranded overhang of the Atto425-labeled conjugate using T4 DNA ligase (NEB). This molecule was purified as described above and then mixed with an excess of streptavidin-coated quantum dots (qd655, Invitrogen) and incubated at room temperature. The qdot-DNA conjugate was purified by repeated ethanol precipitation until few free quantum dots were detected in solution.

All molecules were imaged in TE buffer with 1\% v/v 2-mercaptoethanol added to enhance the fluorescence yield. 
\subsection{Measurements}
\begin{figure}%
\centering
\includegraphics[width=3.25in]{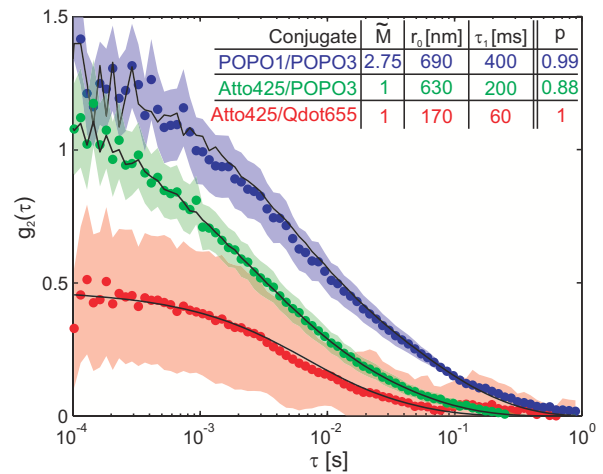}%
\caption{Tracking-FCS data for genomic $\lambda$-phage DNA with varied label configurations. Mean values of $g_2(\tau)$ are indicated by circles, and 2$\sigma$ bounds by the shaded regions. Fit curves to the Rouse model are superimposed in black. The legend indicates the fit parameters and goodness-of-fit p-values as described in the text. The POPO-3/POPO-1 conjugate was prepared with POPO-1 label density of 1 dye per 24000 bp (for an expected $\tilde M \approx 2$ dyes/molecule).}%
\label{Fig:DNAtFCS}%
\end{figure}
Figure \ref{Fig:DNAtFCS} shows $g_2(\tau)$ measured from the three DNA-dye conjugates along with fits to theoretical curves for the Rouse polymer model in which the polymer parameters $\tau_1$ and $r_0$ were fit for all three curves and the mean number $\tilde M$ of intercalating dyes was also fit for the POPO-1/POPO-3 curve. All fits used a first-order tracking system model with $\gamma_{xy} = 15$Hz, $\gamma_z = 2$Hz, $\sigma^0_x = \sigma^0_y = (100\text{nm})^2$ and $\sigma_z^0 = (250\text{nm})^2$ as determined from $\hat D(\tau)$ as described in Section \ref{Section:MSD}. All fits were corrected for attenuation due to background counts\citep{McHale09DNA}. The fits for the POPO-3 labeled molecules include high-frequency triangle-wave oscillations accounting for our excitation scheme, described in \citep{McHale09DNA}, by which we alternatingly expose the molecules to the two excitation lasers to prevent cross-talk between tracking and probe fluorescence. Furthermore, all fits span the range $10^{-4}\text{s} \leq \tau \leq 10^{-1}\text{s}$ due to interference by a systematic decay in our data spanning $0.2\text{s} \leq \tau \leq 1\text{s}$, likely caused by an imperfection in the z-axis tracking\citep{McHale09DNA}. Although this systematic decay is very small compared to the primary tFCS decay, the high statistical resolution on these longer time-scales would place too much emphasis on this spurious component of the tFCS curves if it were included in the fits.

We evaluated the goodness of the fits to the data by assuming that the statistical noise on each of the measured curves is Gaussian, so that the sum of the squared residuals has a $\chi^2$ distribution. The size of the statistical uncertainty in the Atto425/POPO-3 and Atto425/Qdot655 conjugates was determined by computing the variance in $g_2(\tau)$ over the set of observed molecules. The uncertainty in the POPO-1/POPO-3 conjugate was determined using the bootstrap method after processing the data to account for dye configuration uncertainty, as described later in this section.

All three curves fit the data satisfactorily from a statistical perspective, as indicated by the p-values given in the figure. The large p-values for the POPO-3/POPO-1 and Atto425/Qdot655 conjugates suggest that our estimates for the noise in these curves are probably too large.  The consequence of over-estimated uncertainty is reduced ability to discriminate between different models or between different parameter values. Fortunately, we showed in our model discrimination experiments that the difference between the Rouse and Zimm models was great enough to resolve using tFCS on the Atto425/POPO-3 conjugate\citep{McHale09DNA}. 

The POPO-1/POPO-3 and Atto425/POPO-3 conjugates yielded fairly consistent values for $r_0$ and $\tau_1$.  As discussed in \citep{McHale09DNA} these fit parameters, together with the measured translational diffusion statistics (we find $D = 0.8\mu$m$^2/$s), are internally consistent with the predictions from basic polymer theories. By contrast, both fit parameters for the Atto425/Qdot655 conjugate are consistent with those of a molecule that is smaller by a factor of 3. We attribute this difference to non-specific interactions between the DNA and the streptavidin on the qdot surface. Interestingly, such interactions would be very difficult to rule out using stationary FCS because of that technique's insensitivity to molecular size, so that the unfortunate problems with this conjugate highlight a strength of our technique.

While data from the Atto425/POPO-3 and Atto425/Qdot655 conjugates were both straightforward to analyze, the sensitivity of tFCS to the precise positions of dyes in the POPO-1/POPO-3 conjugate demands a more sophisticated approach. For example, Eq. \ref{eq:RouseCorrFcn} predicts both a slower relaxation time and a larger relaxation amplitude for a molecule with a single dye on its end ($u = 0$) compared to a single dye in the middle of the chain ($u = L/2$). Briefly, we account for configuration uncertainty in the POPO-1/POPO-3 conjugate by treating each measured $g_2(\tau)$ as the sum of an ensemble average curve $g^*(\tau)$, a contribution due to dye configuration $\eta(\tau)$, and a statistical noise term $\xi(\tau)$. We estimate the value of $g^*(\tau)$ from a set of measured $g_2(\tau)$ by using a maximum-likelihood estimator; the data in Fig. \ref{Fig:DNAtFCS} is the output of this estimator. Furthermore, since it is difficult to compute $g^*(\tau)$ exactly for the purpose of fitting, we determine the fit curve in the figure by generating 1000 random dye configurations at a fixed dye density and averaging their tFCS curves. Complete details regarding the maximum-likelihood estimator are given in the supporting information. 

\begin{figure}%
\centering
\includegraphics[width=3.25in]{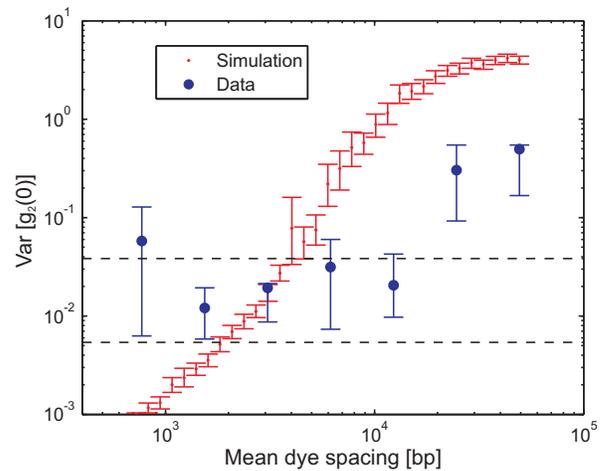}%
\caption{Sensitivity to labeling heterogeneity. The measured variances of $g_2(0)$ within individual DNA samples (blue) are compared to the variances predicted by simulations (red) of ensembles of polymers with uniformly distributed dyes. Dashed lines indicate the shot-noise variances for 15s tracking times with 2.5kHz count rates (upper), and 40s tracking times with 8kHz count rates (lower). Experimental tracking times averaged 25s and count rates averaged 3.8kHz. 95\% error bars were determined by bootstrap sampling.}
\label{Fig:Het}%
\end{figure}

We performed experiments on POPO-1/POPO-3 conjugates over a wide range of POPO-1 densities in order to fully demonstrate the ability of tFCS to resolve differences between molecules based on their dye configurations. During these experiments we varied the probe excitation intensity in order to keep the average count rate, and therefore the measurement noise, roughly constant. The variations between molecules can be evaluated most simply by computing the variance of the measured tFCS curves: expressing the measured curves as a sum of $g^*(\tau)$, $\eta(\tau)$ and $\xi(\tau)$ as described above, we have  $\operatorname{Var}[g_2(\tau)] = \operatorname{Var}[\eta(\tau)] + \operatorname{Var}[\xi(\tau)]$. Since the measurement noise is constant, any differences in $\operatorname{Var}[g_2(\tau)]$ observed in these experiments must derive from the dependence of $\eta(\tau)$ on dye label density. Figure \ref{Fig:Het} shows the results of these measurements. As expected, at sufficiently low dye densities (large dye spacings) the observed variations between molecules exceed the values predicted for measurement noise alone and therefore suggest that we are indeed observing heterogeneity within our sample due to differences in dye configuration.

In order to evaluate the observed heterogeneity quantitatively, we determined the expected variance as a function of dye spacing by using Monte Carlo simulations. These simulations suggest that our observed variances are about a factor of 3 smaller than expected. There are several possible sources of this discrepancy. One is that our prepared dye densities may differ from the true densities, as would be the case for example if some of our DNA were adsorbed to the walls of our sample tubes. However, given that the fitted $\tilde M$ in Fig. \ref{Fig:DNAtFCS} roughly matches the prepared density and that the variances in Fig. \ref{Fig:Het} apparently saturate at similar dye densities, we suspect this is not a major source of error. Another possibility is that our data suffers from selection bias, because we choose to quantitatively examine only fluorescence trajectories for which we clearly see both a POPO-1 and a POPO-3 signal. This selection process is necessary to ensure that dim molecules, which have very poor signal-to-noise and signal-to-background ratios, do not contribute too much noise to our measurements. However, as a consequence of selecting data this way, we discard data from molecules that are either labeled with very few dyes or are labeled more toward their ends than in their centers (because labels near the ends tend to remain further from the molecule's center of mass and, therefore, the probe beam focus). Both exclusions will tend to lower the observed variances, although it is difficult to be certain of the extent to which they have done so in our measurements.
\section{Conclusion}\label{Section:Conclusion}
We derived the FCS statistics measured on large molecules tracked via feedback control. We demonstrated several distinguishing features of these statistics by using a pair of simple molecular models, and demonstrated the application of the tracking-FCS technique to the study of large double-stranded DNA molecules. These results demonstrated our ability to recover realistic and consistent parameters for the dynamics of DNA labeled in two different ways, as well as to identify alterations in the dynamics of the molecule due to attachment of a streptavidin-coated quantum dot label. Furthermore, we demonstrated the ability to resolve configuration differences between different molecules in a sample subjected to a random labeling scheme.  
We have demonstrated that tFCS is a method with unique capabiliites; we believe that these capabilities will establish it as a valuable tool for a range of difficult problems in molecular biophysics.
\section{Acknowledgments}
We thank Aaron Straight and Colin Fuller for suggestions on DNA labeling and purification methods, and Charles Limouse and Michael Zhang for suggestions that improved this manuscript.

\end{document}